\documentclass[preprint,superscriptaddress,preprintnumbers,amsmath,amssymb,prd]{revtex4}
\usepackage{graphicx}

\begin{document}

\thispagestyle{empty}

\title{Quantum field theoretical description of the Casimir effect between
two real graphene sheets and thermodynamics
}

\author{
G.~L.~Klimchitskaya}
\affiliation{Central Astronomical Observatory at Pulkovo of the
Russian Academy of Sciences, Saint Petersburg,
196140, Russia}
\affiliation{Institute of Physics, Nanotechnology and
Telecommunications, Peter the Great Saint Petersburg
Polytechnic University, Saint Petersburg, 195251, Russia}

\author{
V.~M.~Mostepanenko}
\affiliation{Central Astronomical Observatory at Pulkovo of the
Russian Academy of Sciences, Saint Petersburg,
196140, Russia}
\affiliation{Institute of Physics, Nanotechnology and
Telecommunications, Peter the Great Saint Petersburg
Polytechnic University, Saint Petersburg, 195251, Russia}
\affiliation{Kazan Federal University, Kazan, 420008, Russia}

\begin{abstract}
The analytic asymptotic expressions for the Casimir free energy and entropy
for two parallel graphene sheets possessing nonzero energy gap $\Delta$ and
chemical potential $\mu$ are derived at arbitrarily low temperature. Graphene
is described in the framework of thermal quantum field theory in the Matsubara
formulation by means of the polarization tensor in (2+1)-dimensional space-time.
Different asymptotic expressions are found under the conditions $\Delta>2\mu$,
$\Delta=2\mu$, and $\Delta<2\mu$ taking into account both the implicit
temperature dependence due to a summation over the Matsubara frequencies and
the explicit one caused by a dependence of the polarization tensor on
temperature as a parameter. It is shown that for both $\Delta>2\mu$ and
$\Delta<2\mu$ the Casimir entropy satisfies the third law of thermodynamics
(the Nernst heat theorem), whereas for $\Delta=2\mu$ this fundamental
requirement is violated. The physical meaning of the discovered anomaly is
considered in the context of thermodynamic properties of the Casimir effect
between metallic and dielectric bodies.
\end{abstract}

\maketitle
\newcommand{\tp}{{\tilde{\Pi}}}
\newcommand{\rM}{{r_{\rm TM}}}
\newcommand{\rE}{{r_{\rm TE}}}
\newcommand{\orM}{{r_{\rm TM}^{(0)}}}
\newcommand{\orE}{{r_{\rm TE}^{(0)}}}
\newcommand{\zy}{{({\rm i}\zeta_l,y,T)}}
\newcommand{\ozy}{{({\rm i}\zeta_l,y,0)}}
\newcommand{\oyt}{{(0,y,T)}}
\newcommand{\ri}{{{\rm i}}}
\newcommand{\cF}{{\cal{F}}(a,T)}
\newcommand{\ocF}{{\cal{F}}}
\newcommand{\vF}{{\tilde{v}_F}}
\newcommand{\ho}{{\hbar\omega_c}}
\newcommand{\dT}{{\delta_T}}
\newcommand{\daT}{{\delta_T^{\rm impl}}}
\newcommand{\deT}{{\delta_{T}^{\rm expl}}}
\newcommand{\dbT}{{\delta_{T\!,\,l=0}^{\rm expl}\,}}
\newcommand{\dcT}{{\delta_{T\!,\,l\geqslant 1}^{\rm expl}\,}}

\newcommand{\yt}{{(y,T,\Delta,\mu)}}
\newcommand{\yo}{{(y,0,\Delta,\mu)}}
\newcommand{\eEt}{e^{\frac{t\Delta-2\mu}{2k_BT}}}
\newcommand{\deE}{e^{-\frac{\Delta-2\mu}{2k_BT}}}
\newcommand{\meE}{e^{-\frac{2\mu-\Delta}{2k_BT}}}

\section{Introduction}

The Casimir effect was discovered \cite{1} as an attractive force which arises
between two parallel uncharged ideal metal planes in vacuum and depends only
on the Planck constant $\hbar$, speed of light $c$, and interplane distance
$a$. At zero temperature of the planes this effect is entirely caused by the
zero-point oscillations of the quantized electromagnetic field whose spectrum
is altered by the presence of boundary conditions on the planes as compared to
the free Minkowski space. More recently, the Casimir effect was generalized to
the case of metallic or dielectric plates kept at arbitrary temperature $T$.
In the framework of the Lifshitz theory, the free energy and force of the
Casimir interaction between real-material plates are represented as some
functionals of the reflection coefficients expressed via the frequency-dependent
dielectric permittivities of plate materials. Detailed information on
calculation of the Casimir free energies and forces using the Lifshitz theory,
as well as about a comparison between experiment and theory, can be found in
the monograph \cite{2}. There are also generalizations of the Lifshitz theory
for bodies of arbitrary shape and alternative derivations of the Casimir
interaction in the literature (see, e.g., Refs.~\cite{2,2a,2b,2c}).

During the last few years, much attention is given to graphene which is
a one-atom-thick layer of carbon atoms possessing unusual physical
properties \cite{3}. It has been shown that at energies below 1--2~eV
graphene is well described by the Dirac model as a set of massless or very
light electronic quasiparticles. The corresponding fermion field satisfies
the relativistic Dirac equation in (2+1)-dimensions where the speed of
light $c$ is replaced with the Fermi velocity $v_F\approx c$/300 \cite{3,4}.
This allowed application of the methods developed earlier in planar quantum
electrodynamics \cite{5,6,7,8} for investigation of various quantum effects
in graphene systems \cite{9,10,11,12,13,13a,14}.

One of these effects is the Casimir attraction between two parallel
graphene sheets which can be calculated using the Lifshitz theory \cite{2}.
For this purpose, one should know the response function of graphene to the
electromagnetic field which does not reduce to the standard dielectric
permittivities of metallic and dielectric materials. It is important to
keep in mind that the permittivities of ordinary materials are usually
derived using the kinetic theory or Kubo formula under several assumptions
which are not universally applicable \cite{14.0}. These ones and some other
theoretical approaches have been used in approximate calculations of the
response functions and the Casimir force in graphene systems
\cite{14.1,14.2,14.3,14.4,14.5,14.6,14.7,14.8,14.9,14.10,14.11,14.12,14.13,14.14,14.15,14.16,14.17,14.18}. In the framework of the Dirac model, however,
the dielectric response of graphene can be described exactly by means of
its polarization tensor found on the basis of first principles of
thermal quantum field theory.

Although the polarization tensor of graphene was considered in many
papers (see, e.g., Ref.~\cite{15} and literature therein), the exact
expression for it at zero temperature, as well as the corresponding
formulas for the reflection coefficients, have been found in
Ref. \cite{16}. The polarization tensor of gapped graphene (the energy
gap $\Delta$ arises for quasiparticles of nonzero mass) at any temperature
was derived in Ref. \cite{17}. The expressions of Ref. \cite{17} are valid
at the pure imaginary Matsubara frequencies and were used to investigate
the Casimir effect in many graphene systems
\cite{17,18,19,20,21,22,23,24,25,26,27}.
In Ref.~\cite{28} another form for the polarization tensor of
graphene at nonzero temperature was derived valid over the entire plane of
complex frequencies. It was generalized for the case of nonzero chemical
potential $\mu$ in Ref.~\cite{29}. This form of the polarization tensor was
also successfully used in calculations of the Casimir force in various
graphene systems \cite{29,30,31,32,33,34} , as well as for investigation
of the reflectivity and conductivity properties of graphene
\cite{35,36,37,38}.

An interest to the thermodynamic aspects of the Lifshitz theory in
application to graphene systems arose from the so-called Casimir
puzzle. It turned out that the theoretical predictions for the Casimir
force between both metallic and dielectric test bodies are excluded by
the measurement data if one takes into account in calculations the
dissipation of free electrons and the conductivity at a constant current,
respectively (see the reviews in Refs.~\cite{2,39,40} and the most recent
experiments \cite{41,42,43,44}). As to thermodynamics, it was found that
an account of dissipation of free electrons for metals with perfect
crystal lattices and the dc conductivity for dielectrics results in a
violation of the third law of thermodynamics which is also known as the
Nernst heat theorem (see the reviews in Refs.~\cite{2,39} and the most
recent results in Refs.~\cite{45,46,47,48,49,50}). In the single
experiment on measuring the Casimir interaction in graphene systems
performed up to date \cite{51}, the measurement data were found in good
agreement with theoretical predictions using the polarization tensor
\cite{52}. Taking into consideration that the polarization tensor of
graphene results in two spatially nonlocal dielectric permittivities,
the longitudinal one and the transverse one, each of which is
complex and takes dissipation into account, the question arises
whether the Casimir free energy and entropy of graphene systems is
consistent with the requirements of thermodynamics.

To answer this question, the low-temperature behavior of the Casimir
free energy and entropy between two sheets of pristine graphene with
$\Delta=\mu=0$ was found in Ref. \cite{53}. It was shown that in
this case the Casimir entropy vanishes with vanishing temperature, i.e.,
the Nernst heat theorem is satisfied. The same result was obtained for
the Casimir-Polder entropy of an atom interacting with a sheet of a
pristine graphene \cite{54}. For an atom interacting with real graphene
sheet possessing nonzero $\Delta$ and $\mu$ it was shown that the
Nernst heat theorem is followed for $\Delta>2\mu$ \cite{55} and
$\Delta<2\mu$ \cite{55,56}. As to the case $\Delta=2\mu$, the
nonzero value of the Casimir-Polder entropy at zero temperature was
found in this case depending on the parameters of a system, i.e., an
entropic anomaly \cite{56} (the low-temperature behavior of the
Casimir-Polder free energy for $\Delta, \mu\neq$0 was also considered
in Ref.~\cite{57}).

In this paper, we derive the low-temperature analytic asymptotic
expressions for the Casimir free energy and entropy of two real
graphene sheets possessing nonzero values of $\Delta$ and $\mu$.
This is a more complicated problem than for an atom interacting with
real graphene sheet because the free energy of an atom-graphene
interaction is the linear function of the reflection coefficients,
which is not the case for two parallel graphene sheets. The Casimir
free energy is presented by the Lifshitz formula where the reflection
coefficients are expressed via the polarization tensor of graphene
in (2+1)-dimensional space-time. The thermal correction to the Casimir
energy at zero temperature is separated in two contributions. In the
first of them, the temperature dependence is determined exclusively
by a summation over the Matsubara frequencies, whereas the
polarization tensor is defined at zero temperature. The temperature
dependence of the second contribution is determined by an explicit
dependence of the polarization tensor on temperature as a parameter.

We find the asymptotic behaviors at low temperature for each of these
contributions under different relationships between $\Delta$ and 2$\mu$.
It is shown that the leading terms determining the low-temperature
behavior of the total Casimir free energy originate from the first
contribution to the thermal correction for both $\Delta>2\mu$ and
$\Delta<2\mu$ and from the second contribution for $\Delta=2\mu$.
As a result, for $\Delta>2\mu$ and $\Delta<2\mu$ the Nernst heat
theorem is satisfied, whereas for $\Delta=2\mu$ it is violated.
The physical meaning of this anomaly is discussed in the context of
problems considered earlier in the literature on the Casimir effect
between metals and dielectrics.

The paper is organized as follows. In Sec.~II, we briefly summarize
the necessary formalism of the polarization tensor. Section III is
devoted to the perturbation expansion of the Lifshitz formula at low
temperature. In Secs.~IV, V, and VI, the derivation of the asymptotic
expressions for the Casimir free energy and entropy at low temperature
is presented for the cases $\Delta>2\mu$, $\Delta=2\mu$, and
$\Delta<2\mu$, respectively. Section VII contains our conclusions
and a discussion. In the Appendix, the reader will find some
calculation details.

\section{The polarization tensor of graphene and the reflection
coefficients}

We consider two parallel graphene sheets separated by a distance $a$
at temperature $T$ in thermal equilibrium with the environment.
The electronic quasiparticles in graphene considered in the framework
of the Dirac model \cite{3,4} are characterized by some small but nonzero
mass which results in the energy gap $\Delta$ taking the typical value
0.1--0.2~eV. The energy gap arises due to an
impact of the defects of structure, interelectron interactions and
interaction with a substrate if any \cite{15,58}.
We also assume that the graphene sheets under consideration possess
some value of the chemical potential $\mu$ which depends on the doping
concentration \cite{59}  (for a pristine graphene $\Delta=\mu=0$).

The polarization tensor of graphene describes its response to an external
electromagnetic field in the one-loop approximation. The values of this
tensor at the pure imaginary Matsubara frequencies
$\xi_l=2\pi k_BTl/\hbar$ (where $k_B$ is the Boltzmann constant and
$l=0,\,1,\,2\,\ldots$) are usually notated as
\begin{equation}
\Pi_{mn}(\ri\xi_l,k_{\bot},T,\Delta,\mu)\equiv
\Pi_{mn,l}(k_{\bot},T,\Delta,\mu),
\label{eq1}
\end{equation}
\noindent
where $m,\,n=0,\,1,\,2$ are the tensor indices and $k_{\bot}$ is the magnitude
of the wave vector projection on the plane of graphene.
Below it is convenient to consider the dimensionless polarization tensor,
frequencies and the wave vector projection defined as
\begin{equation}
\tp_{mn.l}=\frac{2a}{\hbar}\Pi_{mn,l}, \quad
\zeta_l=\frac{\xi_l}{\omega_c}, \quad
\omega_c\equiv\frac{c}{2a},
\quad
y=2a\left(k_{\bot}^2+\frac{\xi_l^2}{c^2}\right)^{1/2}.
\label{eq2}
\end{equation}

In fact only the two components of the polarization tensor are the independent
quantities. As an example, the 00 component $\tp_{00}$ and the trace
$\tp_{m}^{\,m}$ are often used for a full characterization of this tensor
\cite{17}. For our purposes it is more convenient to use $\tp_{00}$ and
the following linear combination of the 00 component and the trace:
\begin{equation}
\tp_l\equiv\tp(\ri\zeta_l,y,T,\Delta,\mu)
=
(y^2-\zeta_l^2)\tp_m^{\,m}(\ri\zeta_l,y,T,\Delta,\mu)
-y^2\tp_{00}(\ri\zeta_l,y,T,\Delta,\mu).
\label{eq3}
\end{equation}

The reason is that the reflection coefficients on graphene sheets for
the transverse magnetic (TM) and transverse electric (TE) polarizations of
the electromagnetic waves take the following simple form \cite{16,17,28,29}:
\begin{eqnarray}
&&
\rM\zy=\frac{y\tp_{00,l}\yt}{y\tp_{00,l}\yt+2(y^2-\zeta_l^2)},
\nonumber \\[-1mm]
&&\label{eq4}\\[-2mm]
&&
\rE\zy=-\frac{\tp_{l}\yt}{\tp_{l}\yt+2y(y^2-\zeta_l^2)},
\nonumber
\end{eqnarray}
\noindent
where we omitted the parameters $\Delta$ and $\mu$ in the notations of the
reflection coefficients for the sake of brevity.

Now we present the exact expressions for $\tp_{00,l}$ and $\tp_l$ obtained
in the literature. First of all, it is convenient to present them as the
respective quantity defined at $T=0$ plus the thermal correction to it
\begin{eqnarray}
&&
\tp_{00,l}\yt=\tp_{00,l}\yo+\dT\tp_{00,l}\yt,
\nonumber \\
&&
\tp_{l}\yt=\tp_{l}\yo+\dT\tp_{l}\yt.
\label{eq4a}
\end{eqnarray}
\noindent
It is also useful to present $\tp_{00,l}$ and $\tp_l$ as the sums of
contributions which do not depend and, quite the reverse, depend on
$\mu$ and $T$ \cite{34}
\begin{eqnarray}
&&
\tp_{00,l}\yt= \tp_{00,l}^{(0)}(y,\Delta)+ \tp_{00,l}^{(1)}\yt,
\nonumber \\
&&
\tp_{l}\yt= \tp_{l}^{(0)}(y,\Delta)+ \tp_{l}^{(1)}\yt.
\label{eq5}
\end{eqnarray}

As the first contributions on the right-hand side of Eq.~(\ref{eq5}) we choose
the 00 component and the combination (\ref{eq3}) for the polarization tensor
of gapped ($\Delta\neq 0$) but undoped ($\mu=0$) graphene defined at zero
temperature \cite{16,34}
\begin{eqnarray}
&&
\tp_{00,l}^{(0)}(y,\Delta)=\alpha\frac{y^2-\zeta_l^2}{p_l}
\Psi\left(\frac{D}{p_l}\right),
\nonumber \\[-1mm]
&&
\label{eq6}\\[-2mm]
&&
\tp_{l}^{(0)}(y,\Delta)=\alpha(y^2-\zeta_l^2){p_l}
\Psi\left(\frac{D}{p_l}\right),
\nonumber
\end{eqnarray}
\noindent
where $\alpha=e^2/(\hbar c)$ is the fine structure constant,
$D\equiv\Delta/(\ho)$, and the following notations are introduced
\begin{equation}
\Psi(x)=2\left[x+(1-x^2)\arctan(x^{-1})\right],
\quad
p_l=\left[\vF^2y^2+(1-\vF^2)\zeta_l^2\right]^{1/2},
\quad \vF=\frac{v_F}{c}.
\label{eq7}
\end{equation}

In accordance to our choice,
\begin{eqnarray}
&&
\tp_{00,l}^{(0)}(y,\Delta)=\tp_{00,l}(y,0,\Delta,0),
\nonumber \\
&&
\tp_{l}^{(0)}(y,\Delta)=\tp_{l}(y,0,\Delta,0).
\label{eq8}
\end{eqnarray}
\noindent
In so doing, $\tp_{00,l}^{(1)}$ and $\tp_{l}^{(1)}$ acquire
a meaning of the thermal corrections to the
polarization tensor of undoped graphene defined at $T=0$:
\begin{eqnarray}
&&
\tp_{00,l}^{(1)}(y,T,\Delta,0)=\dT\tp_{00,l}(y,T,\Delta,0),
\nonumber \\
&&
\tp_{l}^{(1)}(y,T,\Delta,0)=\dT\tp_{l}(y,T,\Delta,0).
\label{eq9}
\end{eqnarray}
\noindent
These corrections vanish in the limit of zero temperature.

The second contributions on the right-hand side of Eq.~(\ref{eq5}) can be
explicitly presented in the form \cite{34,56}
 \begin{eqnarray}
&&
\tp_{00,l}^{(1)}\yt=\frac{4\alpha D}{\vF^2}\int_1^{\infty}\!\!\!dt
w(t,T,\Delta,\mu)X_{00,l}(t,y,D),
\nonumber\\[-1mm]
&&\label{eq10}\\[-2mm]
&&
\tp_{l}^{(1)}\yt=-\frac{4\alpha D}{\vF^2}\int_1^{\infty}\!\!\!dt
w(t,T,\Delta,\mu)X_{l}(t,y,D),
\nonumber
\end{eqnarray}
\noindent
where the $\mu$-dependent factor is given by
\begin{equation}
w(t,T,\Delta,\mu)=\left(e^{\frac{t\Delta+2\mu}{2k_BT}}+1\right)^{-1}+
\left(e^{\frac{t\Delta-2\mu}{2k_BT}}+1\right)^{-1},
\label{eq11}
\end{equation}
\noindent
and the functions $X_{00,l}$ and $X_l$ are defined as follows:
\begin{widetext}
\begin{eqnarray}
&&
X_{00,l}(t,y,D)=1-{\rm Re}
\frac{p_l^2-D^2t^2+2\ri\zeta_lDt}{\left[p_l^4-p_l^2D^2t^2+\vF^2(y^2-\zeta_l^2)D^2
+2\ri\zeta_lp_l^2Dt\right]^{1/2}},
\nonumber\\[-1mm]
&&\label{eq12}\\[-2mm]
&&
X_{l}(t,y,D)=\zeta_l^2-{\rm Re}
\frac{\zeta_l^2p_l^2-p_l^2D^2t^2+\vF^2(y^2-\zeta_l^2)D^2+
2\ri\zeta_lp_l^2Dt}{\left[p_l^4-p_l^2D^2t^2+\vF^2(y^2-\zeta_l^2)D^2
+2\ri\zeta_lp_l^2Dt\right]^{1/2}}.
\nonumber
\end{eqnarray}
\end{widetext}

It has been shown \cite{33,34} that for a doped and gapped graphene satisfying
the condition $\Delta\geqslant2\mu$ the polarization tensor at $T=0$ also
does not depend on $\mu$. As  a result, one obtains the equalities similar to
those in Eqs.~(\ref{eq8}) and (\ref{eq9})
\begin{equation}
\tp_{00,l}(y,0,\Delta,\mu)=\tp_{00,l}^{(0)}(y,\Delta),
\quad
\tp_{l}(y,0,\Delta,\mu)=\tp_{l}^{(0)}(y,\Delta),
\label{eq13}
\end{equation}
\noindent
and
\begin{equation}
\dT\tp_{00,l}\yt=\tp_{00,l}^{(1)}\yt,
\quad
\dT\tp_{l}\yt=\tp_{l}^{(1)}\yt,
\label{eq14}
\end{equation}
\noindent
where the thermal corrections vanish with vanishing temperature.

It is significant that under the condition $\Delta<2\mu$ the polarization
tensor of doped and gapped graphene at $T=0$ depends both on $\Delta$
and $\mu$, and Eqs.~(\ref {eq13}) and (\ref{eq14}) are not valid any more.
In this case, the 00 component of the polarization tensor at $T=0$ and
the combination of its components (\ref{eq3}) are given by \cite{33}
\begin{widetext}
\begin{eqnarray}
&&
\tp_{00,l}(y,0,\Delta,\mu)=\frac{8\alpha\mu}{\vF^2\ho}-
\frac{2\alpha(y^2-\zeta_l^2)}{p_l^3}\left\{
\vphantom{\left[\frac{\pi}{2}\right]}(p_l^2+D^2)
{\rm Im}\left(z_l\sqrt{1+z_l^2}\right)\right.
\nonumber \\
&&~~~~~~~\left.
+(p_l^2-D^2)\left[
{\rm Im}\ln\left(z_l+\sqrt{1+z_l^2}\right)-\frac{\pi}{2}\right]\right\},
\nonumber \\[-1mm]
&&\label{eq15}\\[-1mm]
&&
\tp_{l}(y,0,\Delta,\mu)=-\frac{8\alpha\mu\zeta_l^2}{\vF^2\ho}+
\frac{2\alpha(y^2-\zeta_l^2)}{p_l}\left\{
\vphantom{\left[\frac{\pi}{2}\right]}(p_l^2+D^2)
{\rm Im}\left(z_l\sqrt{1+z_l^2}\right)\right.
\nonumber \\
&&~~~~~~~\left.
-(p_l^2-D^2)\left[
{\rm Im}\ln\left(z_l+\sqrt{1+z_l^2}\right)-\frac{\pi}{2}\right]\right\},
\nonumber
\end{eqnarray}
\end{widetext}
where
\begin{equation}
z_l\equiv z_l(y,\Delta,\mu)=\frac{p_l}{\vF\sqrt{p_l^2+D^2}\sqrt{y^2-\zeta_l^2}}
\left(\zeta_l+\ri\frac{2\mu}{\ho}\right).
\label{eq16}
\end{equation}

The thermal corrections to the polarization tensor of graphene satisfying the
condition $\Delta<2\mu$ are immediately obtained from Eqs.~(\ref{eq4a})
and (\ref{eq5})
\begin{eqnarray}
&&
\dT\tp_{00,l}\yt=\tp_{00,l}\yt-\tp_{00,l}\yo
\nonumber \\
&&
\phantom{\dT\tp_{00,l}\yt}=\tp_{00,l}^{(1)}\yt-\tp_{00,l}^{(1)}\yo,
\nonumber \\[-1.mm]
&&\label{eq16a}\\[-1.3mm]
&&
\dT\tp_{l}\yt=\tp_{l}\yt-\tp_{l}\yo
\nonumber\\
&&
\phantom{\dT\tp_{l}\yt}=\tp_{l}^{(1)}\yt-\tp_{l}^{(1)}\yo.
\nonumber
\end{eqnarray}

As to the case of an exact equality $\Delta=2\mu$, it is considered in Sec.~V.

\section{Perturbation expansion of the Lifshitz formula at low temperature}

Using the reflection coefficients (\ref{eq4}) expressed above via the
polarization tensor, one can represent the Casimir free energy per unit area
of graphene sheets by means of the Lifshitz formula \cite{2,60}
\begin{equation}
\cF=\frac{k_BT}{8\pi a^2}\sum_{l=0}^{\infty}
{\vphantom{\sum}}^{\prime}
\int_{\zeta_l}^{\infty}\!\!\!\!ydy\sum_{\lambda}
\ln\left[1-r_{\lambda}^2\zy e^{-y}\right],
\label{eq17}
\end{equation}
\noindent
where the prime on the summation sign divides the term with
$l=0$ by 2, and the sum in $\lambda$  is over two polarizations of the
electromagnetic field, transverse magnetic and transverse electric
 ($\lambda={\rm TM,\,TE}$).

We are in fact interested not in the total Casimir free energy but in its
temperature-dependent part, i.e., in the thermal correction to the
Casimir energy defined as
\begin{equation}
\dT\cF=\cF-E(a),
\label{eq18}
\end{equation}
\noindent
where the Casimir energy at zero temperature is given by \cite{2,60}
\begin{equation}
E(a)=\frac{\hbar c}{32\pi^2 a^3}\int_0^{\infty}\!\!d\zeta
\int_{\zeta}^{\infty}\!\!\!\!ydy\sum_{\lambda}
\ln\left[1-r_{\lambda}^2(\ri\zeta,y,0) e^{-y}\right].
\label{eq19}
\end{equation}
\noindent
Here, the reflection coefficients are expressed by Eq.~(\ref{eq4}) in which
one should replace the Matsubara frequencies with a continuous frequency
$\zeta$ and put $T=0$
\begin{eqnarray}
&&
\rM(\ri\zeta,y,0)=
\frac{y\tp_{00}(\ri\zeta,y,0,\Delta,\mu)}{y\tp_{00}(\ri\zeta,y,0,\Delta,\mu)
+2(y^2-\zeta^2)},
\nonumber \\[-1mm]
&&\label{eq20}\\[-2mm]
&&
\rE(\ri\zeta,y,0)=
-\frac{\tp(\ri\zeta,y,0,\Delta,\mu)}{\tp(\ri\zeta,y,0,\Delta,\mu)
+2y(y^2-\zeta^2)}.
\nonumber
\end{eqnarray}
\noindent
Note that both the propagating waves, which are on the mass shell, and the
evanescent waves off the mass shell contribute to Eqs.~(\ref{eq17}) and
(\ref{eq19}).

In the case $\Delta>2\mu$, following Eq.~(\ref{eq13}), one should substitute
to Eq.~(\ref{eq20}) the expressions for the $\tp_{00}$ and $\tp$ defined
in Eq.~(\ref{eq6}) making there the above replacement $\zeta_l\to\zeta$.
If, however, the condition $\Delta<2\mu$ is fulfilled, it is necessary to
substitute in Eq.~(\ref{eq20}) the quantities (\ref{eq15}) with the same
replacement.

Now we identically rearrange Eq.~(\ref{eq18}) to the form
\begin{equation}
\dT\cF=\daT\cF+\deT\cF,
\label{eq21}
\end{equation}
\noindent
where
\begin{widetext}
\begin{eqnarray}
&&
\daT\cF=\frac{k_BT}{8\pi a^2}\sum_{l=0}^{\infty}
{\vphantom{\sum}}^{\prime}
\int_{\zeta_l}^{\infty}\!\!\!\!ydy\sum_{\lambda}
\ln\left[1-r_{\lambda}^2\ozy e^{-y}\right]-E(a)
\label{eq22}\\
&&\hspace*{-2.cm}\mbox{and}
\nonumber\\
&&
\deT\cF=\cF-\frac{k_BT}{8\pi a^2}\sum_{l=0}^{\infty}
{\vphantom{\sum}}^{\prime}
\int_{\zeta_l}^{\infty}\!\!\!\!ydy\sum_{\lambda}
\ln\left[1-r_{\lambda}^2\ozy e^{-y}\right].
\label{eq23}
\end{eqnarray}
\end{widetext}
\noindent
As is seen from Eqs.~(\ref{eq21})--(\ref{eq23}), we have simply added and
subtracted from Eq.~(\ref{eq18}) the quantity having the same form as
the Casimir free energy in Eq.~(\ref{eq17}) but containing the reflection
coefficients (\ref{eq4}) taken at $T=0$.

An advantage of Eq.~(\ref{eq21}) is that the implicit temperature dependence
of the first term, $\daT\ocF$, is entirely determined by a summation on the
Matsubara frequencies, whereas the polarization tensor is taken at $T=0$.
As to the second term, $\deT\ocF$, it simply vanishes for the
temperature-independent polarization tensors. Thus, the dependence of this
term on $T$ can be called explicit.

We turn our attention to the perturbation expansion of the Casimir free energy
at low temperature. Taking into account that the thermal corrections
$\dT\tp_{00,l}$ and $\dT\tp_{l}$ go to zero with vanishing $T$, we substitute
Eq.~(\ref{eq4a}) in Eq.~(\ref{eq4}), expand up to the first order of small
parameters
\begin{equation}
\frac{\dT\tp_{00,l}\yt}{\tp_{00,l}\yo}\ll 1, \quad
\frac{\dT\tp_{l}\yt}{\tp_{l}\yo}\ll 1
\label{eq24}
\end{equation}
\noindent
and obtain
\begin{equation}
r_{\rm TM(TE)}\zy=r_{\rm TM(TE)}\ozy+\dT r_{\rm TM(TE)}\zy,
\label{eq25}
\end{equation}
\noindent
where the first contributions are given by Eq.~(\ref{eq4})
taken at $T=0$ and the thermal corrections to the reflection coefficients
are given by
\begin{eqnarray}
&&
\dT\rM\zy=\frac{2y(y^2-\zeta_l^2)\dT\tp_{00,l}\yt}{[y\tp_{00,l}\yo
+2(y^2-\zeta_l^2)]^2},
\nonumber \\[-0.8mm]
&&\label{eq26}\\[-2mm]
&&
\dT\rE\zy=-\frac{2y(y^2-\zeta_l^2)\dT\tp_{l}\yt}{[\tp_{l}\yo
+2y(y^2-\zeta_l^2)]^2}.
\nonumber
\end{eqnarray}
\noindent
This approach is applicable under the conditions
$\tp_{00,l}\yo\neq 0$ and $\tp_{l}\yo\neq 0$ which are valid
for the cases $\Delta\geqslant 2\mu$ considered in Secs.~IV and V.
For the case $\Delta<2\mu$, however, one cannot use the perturbation
theory in the parameters (\ref{eq24}) for the contribution of the
Matsubara term with $l=0$ (see Sec.~VI).

The implicit thermal correction $\daT\ocF$ defined in Eq.~(\ref{eq22})
is the difference between the sum in $l$ and the integral (\ref{eq19})
with respect to $\zeta$. From Eq.~(\ref{eq2}) it is seen that
$\zeta_l=\tau l$ where $\tau\equiv 4\pi k_BTa/(\hbar c)$.
By replacing the integration variable $\zeta$ in Eq.~(\ref{eq19}) with
$t=\zeta/\tau$, one can bring Eq.~(\ref{eq22})  to the form
\begin{equation}
\daT\cF=\frac{k_BT}{8\pi a^2}\left[
\sum_{l=0}^{\infty}{\vphantom{\sum}}^{\prime}
\Phi(\tau l)-
\int_{0}^{\infty}\!\!\!dt\Phi(\tau t)\right],
\label{eq27}
\end{equation}
\noindent
where
\begin{equation}
\Phi(x)=\int_{x}^{\infty}\!\!\!\!ydy\sum_{\lambda}
\ln\left[1-r_{\lambda}^2(\ri x,y,0) e^{-y}\right].
\label{eq28}
\end{equation}

By applying the Abel-Plana formula \cite{2,60a}, Eq.~(\ref{eq27}) can
be rewritten as
\begin{equation}
\daT\cF=\frac{\ri k_BT}{8\pi a^2}
\int_{0}^{\infty}\!\frac{dt}{e^{2\pi t}-1}\left[
\Phi(\ri \tau t)-\Phi(-\ri \tau t)\right].
\label{eq29}
\end{equation}
\noindent
In the next sections, Eq.~(\ref{eq29}) is used to find the asymptotic
behavior of $\daT\ocF$ at arbitrarily low $T$.

In order to determine the low-temperature behavior of the second thermal
correction to the Casimir energy, $\deT\ocF$, we substitute Eq.~(\ref{eq25})
into its definition (\ref{eq23}) and use the identity
\begin{widetext}
\begin{eqnarray}
&&
\ln\left\{1-\left[r_{\lambda}\ozy+\dT r_{\lambda}\zy\right]^2e^{-y}\right\}
-\ln\left[1-r_{\lambda}^2\ozy e^{-y}\right]
\nonumber \\
&&
=\ln\left\{1-\frac{2r_{\lambda}\ozy\dT r_{\lambda}\zy+
[\dT r_{\lambda}\zy]^2}{1-r_{\lambda}^2\ozy e^{-y}}\,e^{-y}\right\}.
\label{eq30}
\end{eqnarray}
\end{widetext}
\noindent
Then, expanding the logarithm up to the first power of a small parameter and preserving only the term of the first power in $\dT r_{\lambda}\zy$,
one arrives at
\begin{equation}
\deT\cF=-\frac{k_BT}{4\pi a^2}
\sum_{l=0}^{\infty}{\vphantom{\sum}}^{\prime}
\int_{\zeta_l}^{\infty}\!\!\!ydye^{-y}
\sum_{\lambda}
\frac{r_{\lambda}\ozy\dT r_{\lambda}\zy}{1-r_{\lambda}^2\ozy e^{-y}}.
\label{eq31}
\end{equation}
\noindent
This equation valid under a condition that $\tp_{00,l}$ and $\tp_l$ are
nonzero at $T=0$ and, thus, $r_{\lambda}\ozy\neq 0$ is used below to
determine the behavior of $\deT\ocF$ at low temperature.

\section{Low-temperature behavior of the Casimir free energy and entropy for
graphene sheets with {\boldmath$\Delta>2\mu$}}

We assume that the graphene sheets under consideration in this section satisfy the
condition $\Delta>2\mu$ and start with the thermal correction
$\daT\cF$ to the Casimir energy defined in Eq.~(\ref{eq22}) and expressed by
Eqs.~(\ref{eq27}) and (\ref{eq29}). In accordance to Eq.~(\ref{eq28}) the
function $\Phi$ entering Eq.~(\ref{eq27}) is defined as the sum of contributions from the TM and TE modes
\begin{equation}
\Phi(x)=\Phi_{\rm TM}(x)+\Phi_{\rm TE}(x).
\label{eq32}
\end{equation}
\noindent
As a result, $\daT\cF$ becomes the sum of $\daT\ocF_{\rm TM}(a,T)$ and
$\daT\ocF_{\rm TE}(a,T)$.

Under the condition $\Delta>2\mu$, the polarization tensor at $T=0$ is given by
Eq.~(\ref{eq6}). By replacing $\zeta_l$ with $x$ in Eq.~(\ref{eq6}) and
substituting the obtained expressions in Eq.~(\ref{eq20}) where $\zeta$ is also
replaced with $x$, one obtains
\begin{eqnarray}
&&
\rM(\ri x,y,0)=
\frac{\alpha y\Psi(Dp^{-1})}{\alpha y\Psi(Dp^{-1})+2p(x,y)},
\nonumber \\[-1mm]
&&\label{eq33}\\[-2mm]
&&
\rE(\ri x,y,0)=
-\frac{\alpha p(x,y)\Psi(Dp^{-1})}{\alpha p(x,y)\Psi(Dp^{-1})+2y},
\nonumber
\end{eqnarray}
\noindent
where the quantity $p$ is defined as
\begin{equation}
p\equiv p(x,y)=[\vF^2y^2+(1-\vF^2)x^2]^{1/2}.
\label{eq33a}
\end{equation}

In the analytic asymptotic expressions here and below we use the condition
$\Delta>\hbar\omega_c$ (i.e., $D>1$) which is satisfied at not too
small separations between the graphene sheets. Under this condition,
at sufficiently small $x$ (low $T$) one can safely use the inequality
$D\gg p(x,y)$ because the dominant contribution to the integrals in
Eq.~(\ref{eq28}) is given by $y\sim 1$.

We consider first the case $\lambda={\rm TM}$. By expanding in Eq.~(\ref{eq28})
in Taylor series around $x_0=0$ with the help of the first formula in
Eq.~(\ref{eq33}) and above condition, we find
\begin{eqnarray}
&&
\Phi_{\rm TM}(x)=\Phi_{\rm TM}(0)+\frac{4\alpha^2}{9D^2}x^4+
\frac{16\alpha^2(8\alpha+3D)}{135D^3}x^5+O(x^6)
\nonumber \\
&&
~~~~\approx
\Phi_{\rm TM}(0)+\frac{4\alpha^2}{9D^2}x^4+
\frac{16\alpha^2}{45D^2}x^5+O(x^6).
\label{eq34}
\end{eqnarray}

The first two terms on the right-hand side of this equation  do not contribute
to Eq.~(\ref{eq29}), whereas the third term leads to
\begin{equation}
\Phi_{\rm TM}(\ri\tau t)-\Phi_{\rm TM}(-\ri\tau t)=
\ri\frac{32\alpha^2}{45D^2}\tau^5t^5.
\label{eq35}
\end{equation}
\noindent
Substituting this result in Eq.~(\ref{eq29}), one arrives at
\begin{equation}
\daT\ocF_{\rm TM}(a,T)=
-\frac{16\alpha^2\pi^4a(k_BT)^6}{315\Delta^2(\hbar c)^3}.
\label{eq36}
\end{equation}

We continue with the case $\lambda={\rm TE}$. The function $\Phi_{\rm TE}(x)$
cannot be expanded in Taylor series around the point $x_0=0$. Because of this,
we substitute the second line of Eq.~(\ref{eq33}) in Eq.~(\ref{eq28}),
expand the integrand in powers of $x$ and integrate with respect to $y$
thereafter. The result is
\begin{eqnarray}
&&
\Phi_{\rm TE}(x)=\left(\frac{4\alpha}{3D}\right)^2\left[
\vphantom{\left(1-\frac{3}{4}\vF^2\right)}
-6\vF^4-2\vF^2(1-\vF^2)x^2
\right.
+\vF^2\left(1-\frac{3}{4}\vF^2\right)x^4+(1-\vF^2)x^4{\rm Ei}(-x)
\nonumber \\
&&~~~~~~~~
\left.
-\frac{2\vF^2}{3}
\left(1-\frac{7}{10}\vF^2\right)x^5+O(x^6)\right],
\label{eq37}
\end{eqnarray}
\noindent
where ${\rm Ei}(z)$ in the exponential integral.

The first three terms on the right-hand side of this expression do not contribute
to Eq.~(\ref{eq29}). The dominant contribution is given by the term containing
the exponential integral which leads to
\begin{equation}
\Phi_{\rm TE}(\ri\tau t)-\Phi_{\rm TE}(-\ri\tau t)=
\ri\pi\left(\frac{4\alpha}{3D}\right)^2 \tau^4t^4.
\label{eq38}
\end{equation}
\noindent
Substituting this equation in Eq.~(\ref{eq29}) and integrating, one arrives at
the result
\begin{equation}
\daT\ocF_{\rm TE}(a,T)=
-\frac{32\zeta(5)\alpha^2(k_BT)^5}{3\pi^2\Delta^2(\hbar c)^2}.
\label{eq39}
\end{equation}

Comparing this with Eq.~(\ref{eq36}), we conclude that the dominant term in the
asymptotic behavior of $\daT\ocF$ at low $T$ is given by Eq.~(\ref{eq39}) and
determined by the contribution of the TE mode, i.e.,
\begin{equation}
\daT\cF=\daT\ocF_{\rm TE}(a,T)\sim
-\frac{\alpha^2(k_BT)^5}{\Delta^2(\hbar c)^2}.
\label{eq39a}
\end{equation}

We are now coming to the asymptotic behavior of the second thermal correction,
$\deT\ocF$, at low $T$ which takes into account an explicit dependence of the
polarization tensor on temperature as a parameter. This correction is presented
in Eq.~(\ref{eq31}). It is convenient to express  $\deT\ocF$ as a sum
of two contributions
\begin{equation}
\deT\cF=\dbT\cF+\dcT\cF,
\label{eq40}
\end{equation}
\noindent
where the first one contains the term of Eq.~(\ref{eq31}) with $l=0$ and
the second one --- all terms with $l\geqslant 1$.

We start from the first contribution on the right-hand side of Eq.~(\ref{eq40}).
According to Eq.~(\ref{eq31}), it contains the zero-temperature reflection
coefficients and thermal corrections to them, both taken at the zero Matsubara
frequency. The reflection coefficients at $l=0$ are obtained from Eq.~(\ref{eq33})
by putting $x=0$
\begin{eqnarray}
&&
\rM(0,y,0)=
\frac{\alpha \Psi(D\vF^{-1}y^{-1})}{\alpha \Psi(D\vF^{-1}y^{-1})+2\vF},
\nonumber \\[-1mm]
&&\label{eq41}\\[-2mm]
&&
\rE(0,y,0)=
-\frac{\alpha \vF\Psi(D\vF^{-1}y^{-1})}{\alpha \vF\Psi(D\vF^{-1}y^{-1})+2},
\nonumber
\end{eqnarray}

Taking into account that for $y\sim 1$ it holds $\vF y\ll D$, we expand the
function $\Psi$ in powers of the small parameter $\vF y/D$ and obtain
\begin{equation}
\Psi(D\vF^{-1}y^{-1}) \approx\frac{8}{3}\,\frac{\vF y}{D}.
\label{eq42}
\end{equation}
\noindent
As a result, Eq.~(\ref{eq41}) reduces to
\begin{eqnarray}
&&
\rM(0,y,0)\approx
\frac{\alpha y}{\alpha y+\frac{3}{4}D} \approx
\frac{4\alpha y}{{3}D},
\nonumber \\[-1mm]
&&\label{eq43}\\[-2mm]
&&
\rE(0,y,0)\approx
-\frac{\alpha \vF^2 y}{\alpha \vF^2 y+\frac{3}{4}D} \approx
-\frac{4\alpha \vF^2 y}{{3}D}.
\nonumber
\end{eqnarray}
\noindent
{}From Eq.~(\ref{eq43}) it is seen that
\begin{equation}
\rE(0,y,0)\approx -\vF^2 \rM(0,y,0),
\label{eq44}
\end{equation}
\noindent
i.e., the magnitude of the TE reflection coefficient taken at zero frequency
and temperature is negligibly small as compared to the TM one.

Next, we consider the thermal corrections to the reflection coefficients
(\ref{eq43}) entering Eq.~(\ref{eq31}). By putting $l=0$ in Eq.~(\ref{eq26}),
one obtains
\begin{eqnarray}
&&
\dT\rM(0,y,T)=\frac{2y\dT\tp_{00,0}\yt}{[\tp_{00,0}\yo
+2y]^2},
\nonumber \\[-0.8mm]
&&\label{eq45}\\[-2mm]
&&
\dT\rE(0,y,T)=-\frac{2y^3\dT\tp_{0}\yt}{[\tp_{0}\yo
+2y^3]^2}.
\nonumber
\end{eqnarray}
\noindent
Under the condition $\Delta>2\mu$ we can use Eq.~(\ref{eq14}) and, thus,
the quantities $\dT\tp_{00,0}$ and $\dT\tp_0$ can be obtained from
Eq.~(\ref{eq10}) taken at $l=0$. Taking into account that under the
condition $\Delta>2\mu$ the first contribution to Eq.~(\ref{eq11})
leads to
an additional exponentially small factor $\exp[-2\mu/(k_BT)]$, one can
preserve only the second contribution. As a result, we have
\begin{equation}
\dT\tp_{00,0}\yt=\frac{4\alpha D}{\vF^2}\left[I_{00,0}^{(1)}+
\frac{1}{\vF y}I_{00,0}^{(2)}\right],
\label{eq46}
\end{equation}
\noindent
where
\begin{eqnarray}
&&
I_{00,0}^{(1)}=\int_1^{\infty}\!\!dt
\left(e^{\frac{t\Delta-2\mu}{2k_BT}}+1\right)^{-1},
\label{eq47} \\
&&
I_{00,0}^{(2)}=\int_1^{f(y,D)}\!\!dt
\left(e^{\frac{t\Delta-2\mu}{2k_BT}}+1\right)^{-1}\!\!
\frac{D^2t^2-\vF^2y^2}{[\vF^2y^2-D^2(t^2-1)]^{1/2}}
\nonumber
\end{eqnarray}
\noindent
and the function $f(y,D)$ is defined as
\begin{equation}
f(y,D)=\sqrt{1+\frac{\vF^2y^2}{D^2}}.
\label{eq48}
\end{equation}

For the thermal correction $\dT\tp_0$ from the second line in Eq.~(\ref{eq10})
one obtains
\begin{equation}
\dT\tp_{0}\yt=-\frac{4\alpha D^3y}{\vF}\int_1^{f(y,D)}\!\!dt
\left(e^{\frac{t\Delta-2\mu}{2k_BT}}+1\right)^{-1}\!
\frac{t^2-1}{[\vF^2y^2-D^2(t^2-1)]^{1/2}}.
\label{eq49}
\end{equation}

Since we consider arbitrarily low $T$, we can use the condition
$\Delta-2\mu\gg k_BT$. Under this condition the quantity $I_{00,0}^{(1)}$ in
Eq.~(\ref{eq47}) takes an especially simple form
\begin{equation}
I_{00,0}^{(1)}\approx \frac{2k_BT}{\Delta}\,e^{-\frac{\Delta-2\mu}{2k_BT}}.
\label{eq50}
\end{equation}.

The quantity $I_{00,0}^{(2)}$ defined in Eq.~(\ref{eq47}) is calculated at low
temperature in the Appendix. According to Eq.~(\ref{A3}), the asymptotic
behavior of $I_{00,0}^{(2)}$ is given by
\begin{equation}
I_{00,0}^{(2)}\sim \ \frac{k_BT}{\vF}\frac{\Delta}{(\hbar \omega_c)^2}
\deE.
\label{eq51}
\end{equation}

Then, from Eqs.~(\ref{eq46}), (\ref{eq50}), and (\ref{eq51}) we can conclude
that
\begin{equation}
\dT\tp_{00,0}\yt\sim \frac{\alpha k_BT}{\hbar \omega_c}\deE
\left(C_1+\frac{C_2}{y}\right),
\label{eq54}
\end{equation}
\noindent
where $C_1\sim \vF^{-2}$ and $C_2\sim \vF^{-4}$ are the constants.

The integral with respect to $t$ in Eq.~(\ref{eq49}) for $\dT\tp_0$  can be
estimated similar to Eqs.~(\ref{A2}) and (\ref{A3}). For this purpose,
using Eq.~(\ref{eq48}), we replace $t^2-1$ with a larger quantity
$\vF^2 y^2/D^2$ and obtain
\begin{equation}
\dT\tp_{0}\yt\sim -\frac{\alpha k_BT}{\hbar \omega_c}C_3\deE,
\label{eq55}
\end{equation}
\noindent
where $C_3\sim \vF^{0}$.

Substituting Eqs.~(\ref{eq6}), (\ref{eq42}), (\ref{eq54}) and (\ref{eq55})
in Eq.~(\ref{eq45}), one finds
\begin{eqnarray}
&&\dT\rM\oyt=\frac{\dT\tp_{00,0}\yt}{2y\left(\alpha y\frac{4}{3D}+1\right)^2}
\approx \frac{\dT\tp_{00,0}\yt}{2y}
\sim\frac{\alpha k_BT}{\hbar \omega_c}\deE
\left(\frac{C_1}{y}+\frac{C_2}{y^2}\right),
\nonumber \\
&&\dT\rE\oyt=-\frac{\dT\tp_{0}\yt}{2y^3\left(\alpha \vF^2\frac{4y}{3D}+1\right)^2}
\approx -\frac{\dT\tp_{0}\yt}{2y^3}
\sim\frac{\alpha k_BT}{\hbar \omega_c y^3}C_3\deE.
\label{eq56}
\end{eqnarray}

{}From these equations, we obtain
\begin{equation}
\dT\rE\oyt\sim\vF^4\dT\rM\oyt,
\label{eq57}
\end{equation}
\noindent
i.e., similar to Eq.~(\ref{eq44}), thermal correction to the TE reflection
coefficient at zero Matsubara frequency is negligibly small comparing to
the TM one.

Now we substitute the first lines of Eqs.~(\ref{eq43}) and (\ref{eq56})
in the term of Eq.~(\ref{eq31}) with $l=0$ and obtain
\begin{equation}
\dbT\cF\approx\dbT\ocF_{\rm TM}(a,T)\sim
-\frac{\alpha^2(k_BT)^2}{a^2\Delta}
\deE\int_0^{\infty}\!\!\!dy\,e^{-y}
\frac{C_1y+C_2}{1-\left(\frac{4\alpha y}{3D}\right)^2e^{-y}}.
\label{eq58}
\end{equation}

Taking into consideration that the integral in this equation converges,
the final result is
\begin{equation}
\dbT\cF\sim
-\frac{\alpha^2(k_BT)^2}{a^2\Delta}\deE
\label{eq59}
\end{equation}

We are passing now to a consideration of the correction $\dcT\ocF$ which is
equal to the sum of all terms with $l\geqslant 1$ in Eq.~(\ref{eq31}).
In this case, from Eq.~(\ref{eq33}) with $x=\zeta_l$, using an approximate
equality
\begin{equation}
\Psi\left(\frac{D}{p_l}\right) \approx\frac{8}{3}\,\frac{p_l}{D}
\label{eq60}
\end{equation}
\noindent
similar to Eq.~(\ref{eq42}), we find
\begin{eqnarray}
&&
\rM\ozy\approx
\frac{\alpha y}{\alpha y+\frac{3}{4}D} \approx
\frac{4\alpha y}{{3}D},
\label{eq61}\\
&&
\rE\ozy\approx
-\frac{\alpha p_l^2}{\alpha p_l^2+\frac{3}{4}Dy} \approx
-\frac{4\alpha p_l^2}{{3}Dy}\approx
-\frac{4\alpha \vF^2 y}{{3}D}.
\nonumber
\end{eqnarray}
\noindent
Here we have used that for $y\sim 1$, giving the dominant contribution
to Eq.~(\ref{eq31}), $D\gg\alpha y$ and considered $p_l\approx\vF y$ at
$\tau\to 0$. From Eq.~(\ref{eq61}) it is seen that similar to Eq.~(\ref{eq44})
relationship
\begin{equation}
\rE(\ri\zeta_l,y,0)\approx -\vF^2 \rM(\ri\zeta_l,y,0),
\label{eq62}
\end{equation}
\noindent
holds at any $\zeta_l$.

Using Eq.~(\ref{eq26}), in the same approximation as in Eq.~(\ref{eq56})
one obtains
\begin{eqnarray}
&&\dT\rM\zy\approx \frac{y\dT\tp_{00,l}\yt}{2(y^2-\zeta_l^2)},
\nonumber \\
&&\dT\rE\zy
\approx -\frac{\dT\tp_{l}\yt}{2y(y^2-\zeta_l^2)}.
\label{eq63}
\end{eqnarray}

{}From Eqs.~(\ref{eq10}), (\ref{eq12}) and (\ref{eq14}) one can make sure that
\begin{equation}
\left.\dT\tp_{00,l}\yt\right|_{y=\zeta_l}\!\!\!=
\left.\dT\tp_{l}\yt\right|_{y=\zeta_l\!\!\!}=0.
\label{eq64}
\end{equation}
\noindent
Because of this, the integrals with respect to $y$ in Eq.~(\ref{eq31}) are
convergent at the low integration limit for all $l\geqslant 1$.
Since the dominant contribution in Eq.~(\ref{eq31}) is given by $y\sim 1$,
in the limiting case $\tau\to 0$ one can expand the integrand in Taylor
series in the powers of $\zeta_l=\tau l$.
For the order of magnitude estimation of the asymptotic behavior at $T\to 0$,
it will suffice to consider the lowest expansion order. In this way, from
Eqs.~(\ref{eq31}), (\ref{eq54}) and (\ref{eq63}) we find
\begin{widetext}
\begin{eqnarray}
&&
\dcT\ocF_{\rm TM}(a,T)\sim -\frac{k_BT}{a^2}\sum_{l=1}^{\infty}
\int_{\zeta_l}^{\infty}\!\!\!\!ydye^{-y}
\frac{\rM(0,y,0)}{1-r_{\rm TM}^2(0,y,0)e^{-y}}\frac{\dT\tp_{00,0}\yt}{y}
\nonumber \\
&&~~
\sim -\frac{\alpha(k_BT)^2}{\hbar c a}\deE\sum_{l=1}^{\infty}
\int_{\zeta_l}^{\infty}\!\!\!\!dye^{-y}
\frac{\rM(0,y,0)}{1-r^2_{\rm TM}(0,y,0)e^{-y}}\left(C_1+\frac{C_2}{y}\right).
\label{eq65}
\end{eqnarray}
\end{widetext}
\noindent
By introducing the variable $v=y/\zeta_l$ and using Eq.~(\ref{eq61}), it is seen that in the asymptotic limit $\tau\to 0$ the denominator
in Eq.~(\ref{eq65}) can
be replaced with unity and, thus,
\begin{widetext}
\begin{eqnarray}
&&
\dcT\ocF_{\rm TM}(a,T)\sim -\frac{\alpha^2(k_BT)^2}{\hbar c a}\deE
\sum_{l=1}^{\infty}\zeta_l^2
\int_{\zeta_l}^{\infty}\!\!\!\!vdve^{-\zeta_lv}
\left(C_1+\frac{C_2}{\zeta_lv}\right)
\label{eq66} \\
&&~~
= -\frac{\alpha^2(k_BT)^2}{\hbar c a}\deE\sum_{l=1}^{\infty}
\left[C_1(1+\zeta_l)+C_2\right]e^{-\zeta_l}
\sim -\frac{\alpha^2(k_BT)^2}{\hbar c a}\deE\frac{1}{\tau}
\sim -\frac{\alpha^2k_BT}{a^2}\deE\, .
\nonumber
\end{eqnarray}
\end{widetext}

Similar estimation shows that the contribution of the TE mode to
Eq.~(\ref{eq31}) is again negligibly small
\begin{equation}
\dcT\ocF_{\rm TE}(a,T)\sim \vF^2\dcT\ocF_{\rm TM}(a,T).
\label{eq67}
\end{equation}
\noindent
Because of this, the result is
\begin{equation}
\dcT\cF\sim \dcT\ocF_{\rm TM}(a,T)\sim  -\frac{\alpha^2k_BT}{a^2}\deE.
\label{eq68}
\end{equation}

Comparing Eqs.~(\ref{eq59}) and (\ref{eq68}), we notice that a summation over
the nonzero Matsubara frequencies decreases by one the power of
temperature in
front of the main exponential factor. Note also that Eqs.~(\ref{eq39a}),
(\ref{eq59}), and (\ref{eq68}) are obtained under the condition
$\Delta>\hbar\omega_c$ and, thus, one cannot put there $\Delta=0$.
These equations, however, are well applicable for graphene
with $\mu=0$.

Now we can find the dominant asymptotic behavior of the total thermal correction to
the Casimir energy at zero temperature $\dT\ocF$ in the limit of low temperature.
Taking into account that in accordance to Eqs.~(\ref{eq21}) and (\ref{eq40})
$\dT\ocF$ is given by the sum of Eqs.~(\ref{eq39a}), (\ref{eq59}), and
(\ref{eq68}), one concludes that under a condition $\Delta>2\mu$ its leading
behavior is given by Eq.~(\ref{eq39a}), i.e.,
\begin{equation}
\dT\cF\sim
-\frac{\alpha^2(k_BT)^5}{\Delta^2(\hbar c)^2},
\label{eq69}
\end{equation}
\noindent
and is determined by the TE contribution to the implicit temperature
dependence.

This result gives the possibility to find the low-temperature behavior of the
Casimir entropy per unit area of the graphene sheets defined as
\begin{equation}
S(a,T)=-\frac{\partial\cF}{\partial T}=-\frac{\partial\dT\cF}{\partial T}.
\label{eq70}
\end{equation}
\noindent

Using Eq.~(\ref{eq69}), one finds
\begin{equation}
S(a,T)\sim
\frac{\alpha^2k_B^5T^4}{\Delta^2(\hbar c)^2},
\label{eq71}
\end{equation}
\noindent
which vanishes with vanishing temperature in agreement with the third law of
thermodynamics (the Nernst heat theorem) \cite{61,62}. This means that
the Lifshitz theory using the response function of graphene with $\Delta>2\mu$
expressed in terms of the polarization tensor is thermodynamically consistent.

To summarize the application region of the obtained results, in this section
we used the conditions
\begin{equation}
k_BT\ll\frac{\hbar v_F}{2a}\ll\frac{\hbar c}{2a}<\Delta,
\quad
k_BT\ll\Delta-2\mu
\label{eq73a}
\end{equation}
\noindent
and made the asymptotic expansions in three small parameters
\begin{equation}
\tau\equiv\frac{4\pi k_BTa}{\hbar c}\ll 1, \quad
\frac{\hbar v_F}{2a\Delta}\ll 1, \quad
e^{-\frac{\Delta-2\mu}{2k_BT}}\ll 1.
\label{eq73b}
\end{equation}
\noindent
The last parameter was used in finding the low-temperature behavior of
$\deT{\cal F}$. It is possible, however, to dispense with this parameter
(see the next section).


\section{Low-temperature behavior of the Casimir free energy and entropy for
graphene sheets with {\boldmath$\Delta=2\mu$}}

As was stated in Sec.~II, Eqs.~(\ref{eq13}) and (\ref{eq14}) preserve their
validity in the case $\Delta=2\mu$. Because of this, all the results for
$\daT\ocF$ obtained in Sec.~III for the graphene sheets with $\Delta>2\mu$
remain valid in the case $\Delta=2\mu$. Specifically, the low-temperature
behavior of $\daT\ocF$ is again determined by the TE mode and is given by
Eq.~(\ref{eq39a}).

An explicit temperature dependence, however, leads to a radically different
results. Although Eqs.~(\ref{eq40})--(\ref{eq49}) remain valid in the case
$\Delta=2\mu$, the subsequent equations obtained under a condition
$\Delta-2\mu\gg k_BT$ are not applicable. Thus, instead of Eq.~(\ref{eq50}),
from the first line of Eq.~(\ref{eq47}) we obtain
\begin{equation}
I_{00,0}^{(1)}=\frac{2k_BT}{\Delta}\ln 2.
\label{eq72}
\end{equation}

A more exact calculation of the integral $I_{00,0}^{(2)}$ defined in
Eqs.~(\ref{eq47}) and (\ref{eq48}) in the case $\Delta=2\mu$ (see Appendix)
in accordance to Eq.~(\ref{A6}) results in
\begin{equation}
I_{00,0}^{(2)}\sim\frac{k_BT}{\vF}\frac{\Delta}{(\hbar\omega_c)^2}\ln 2.
\label{eq73}
\end{equation}

As is seen from the comparison of Eqs.~(\ref{eq72}) and (\ref{eq73}) with
Eqs.~(\ref{eq50}) and (\ref{eq51}), respectively, the values of
$I_{00,0}^{(1)}$ and $I_{00,0}^{(2)}$ in the cases $\Delta>2\mu$ and
$\Delta=2\mu$ differ only by the missing exponential factor and by
an occurrence of the factor $\ln 2$ in the latter case. This allows to
conclude that, similar to the case  $\Delta>2\mu$ considered in Sec.~IV,
the dominant contribution to the thermal correction $\dbT\ocF$ is
determined by the TM mode. Up to an order of magnitude estimation of
this contribution for the case $\Delta=2\mu$, in accordance to Eq.~(\ref{eq59}),
is given by
\begin{equation}
\dbT\cF\sim
-\frac{\alpha^2(k_BT)^2}{a^2\Delta}.
\label{eq76}
\end{equation}

In a similar way, by repeating the derivation in Eqs.~(\ref{eq60})--(\ref{eq68}),
one arrives at a conclusion that for $\Delta=2\mu$ the contribution
$\dcT\ocF$ to the thermal correction at low temperature is estimated by   Eq.~(\ref{eq68}) where the exponential factor is replaced with unity
\begin{equation}
\dcT\cF\sim
-\frac{\alpha^2k_BT}{a^2}.
\label{eq77}
\end{equation}

{}From the comparison of Eq.~(\ref{eq39a}) for an implicit contribution
to the thermal correction, which is valid also for the case $\Delta=2\mu$,
with the explicit contributions (\ref{eq76}) and (\ref{eq77}), one concludes
that in this case the low-temperature behavior of the total thermal correction
is given by
\begin{equation}
\dT\cF\sim
-\frac{\alpha^2k_BT}{a^2},
\label{eq78}
\end{equation}
\noindent
which originates from the TM mode in an explicit temperature dependence.
In the case $\Delta=2\mu$, Eqs.~(\ref{eq39a}) and (\ref{eq76})--(\ref{eq78})
are obtained under the first set of inequalities in Eq.~(\ref{eq73a}), i.e.,
do not using the condition $k_BT\ll\Delta-2\mu$.
They employ only the first two small parameters indicated in Eq.~(\ref{eq73b})
and are valid for graphene with $\Delta\neq 0$ and $\mu\neq 0$.

The result (\ref{eq78}) leads to problems. The point is that, in accordance
to Eq.~(\ref{eq70}), the respective Casimir entropy per unit area of the
graphene sheets at low temperature behaves as
\begin{equation}
S(a,T)\sim
\frac{\alpha^2k_B}{a^2}.
\label{eq79}
\end{equation}

Thus, the Casimir entropy at zero temperature is the nonzero (positive)
constant depending on the volume of a system in violation of the Nernst
heat theorem \cite{61,62}. As discussed in Sec.~I, the same situation
holds for metals with perfect crystal lattices described by the dielectric
permittivity of the Drude model which, as opposed to the polarization
tensor of graphene, is not derived from the first principles of quantum
field theory. It should be taken into consideration, however, that for
a real graphene sheet the values of $\Delta$ and $\mu$ cannot be known
precisely. Thus, from the practical standpoint, the equality  $\Delta=2\mu$
can be considered as some singular point (see further discussion in Sec.~VII).
It is only important what are
the properties of the Casimir free energy and entropy at low temperatures
for graphene sheets with $\Delta<2\mu$. This question is answered in the
next section.

\section{Low-temperature behavior of the Casimir free energy and entropy for
graphene sheets with {\boldmath$\Delta<2\mu$}}

Here, we consider the last possibility when the chemical potential is
relatively large by exceeding the half of the energy gap. As in two
preceding sections, we begin with consideration of the implicit contribution
to the thermal correction given by Eq.~(\ref{eq29}),
where the function $\Phi(x)$
is expressed via the reflection coefficients at zero temperature
by Eq.~(\ref{eq28}).

In order to find these reflection coefficients, we consider the polarization
tensor (\ref{eq15}) and (\ref{eq16}) found in the case $\Delta<2\mu$,
replace $\zeta_l$ with $x$ in Eqs.~(\ref{eq15}) and (\ref{eq16}) and expand
the results up to the first power in $x$ under the condition
$\sqrt{4\mu^2-\Delta^2}>\hbar\omega_c$ which is satisfied at not too small
separations between the graphene sheets.
The result is
\begin{equation}
\tp_{00}(x,y,0,\Delta,\mu)=Q_0-Q_1\frac{x}{y},
\quad
\tp(x,y,0,\Delta,\mu)=Q_2yx,
\label{eq80}
\end{equation}
\noindent
where the following notations are introduced
\begin{equation}
Q_0=\frac{4\alpha}{\vF^2}\,\frac{2\mu}{\hbar\omega_c},
\quad
Q_1=\frac{16\alpha\mu^2}{\vF^3\hbar\omega_c\sqrt{4\mu^2-\Delta^2}},
\quad
Q_2=\frac{4\alpha\sqrt{4\mu^2-\Delta^2}}{\vF\hbar\omega_c}.
\label{eq81}
\end{equation}
\noindent
It is easily seen that under the used conditions $Q_0\gg 1$ holds.

We consider first the TM contribution to the function $\Phi(x)$ in
Eqs.~(\ref{eq28}) and (\ref{eq32}) and expand it up to the first power
in small $x$
\begin{equation}
\Phi_{\rm TM}(x)=\Phi_{\rm TM}(0)+x\Phi_{\rm TM}^{\prime}(0).
\label{eq82}
\end{equation}

Substituting  Eq.~(\ref{eq80}) in the first line of Eq.~(\ref{eq20}),
where $\zeta$ is replaced with $x$, one obtains
\begin{eqnarray}
&&
\rM(x,y,0)=\frac{yQ_0-Q_1x}{ yQ_0-Q_1x+2(y^2-x^2)},
\nonumber \\
&&
\rM(0,y,0)=\frac{Q_0}{ Q_0+2y}.
\label{eq83}
\end{eqnarray}

{}From Eq.~(\ref{eq28}) at $\lambda={\rm TM}$, using Eq.~(\ref{eq83}), it is
easily seen that the quantity $\Phi_{\rm TM}(x)$ at $x=0$ is represented by
a converging integral. Calculating the first derivative of
$\Phi_{\rm TM}(x)$, one obtains
\begin{equation}
\Phi_{\rm TM}^{\prime}(x)=-x\ln(1-e^{-x})
-\int_x^{\infty}\!\!\!ydy\frac{2\rM(x,y,0)e^{-y}}{1-r_{\rm TM}^2(x,y,0)e^{-y}}
\frac{\partial\rM(x,y,0)}{\partial x}.
\label{eq84}
\end{equation}

By differentiating the first equality in Eq.~(\ref{eq83}), one finds
\begin{equation}
\left.\frac{\partial\rM(x,y,0)}{\partial x}\right|_{x=0}=
-\frac{2Q_1}{(Q_0+2y)^2}.
\label{eq85}
\end{equation}
\noindent
Then, substituting Eq.~(\ref{eq85}) in Eq.~(\ref{eq84}), we have
\begin{equation}
\Phi_{\rm TM}^{\prime}(0)=4Q_1
\int_0^{\infty}\!\!\!dy\frac{y}{(Q_0+2y)^2}
\frac{\rM(0,y,0)e^{-y}}{1-r_{\rm TM}^2(0,y,0)e^{-y}}.
\label{eq86}
\end{equation}

Taking into account that $Q_0\gg 1$ and that the main contribution to the
integral is given by $y\sim 1$, one finds from the second equality
in Eq.~(\ref{eq83}) that $\rM(0,y,0)\approx 1$. In such a manner,
Eq.~(\ref{eq86}) reduces to
\begin{equation}
\Phi_{\rm TM}^{\prime}(0)\approx\frac{4Q_1}{Q_0^2}
\int_0^{\infty}\!\!\frac{y\,dy}{e^y-1}
=\frac{2\pi^2Q_1}{3Q_0^2}.
\label{eq87}
\end{equation}
\noindent
Substituting this equation in Eq.~(\ref{eq82}), one obtains
\begin{equation}
\Phi_{\rm TM}(\ri\tau t)-\Phi_{\rm TM}(-\ri\tau t)=
\ri\frac{4\pi^2Q_1}{3Q_0^2}\tau T.
\label{eq88}
\end{equation}

Now we consider the contribution of the TE mode in
Eqs.~(\ref{eq28}) and (\ref{eq32}).
In this case the reflection coefficient is obtained by substituting
 Eq.~(\ref{eq80}) in the second line of Eq.~(\ref{eq20})
\begin{equation}
\rE(x,y,0)=-\frac{Q_2x}{Q_2x+2(y^2-x^2)}.
\label{eq89}
\end{equation}
\noindent
As is seen from this equation, $\rE(x,y,0)$ goes to zero with vanishing $x$.

Using the first expansion term in the powers of $\rE(x,y,0)$ in Eq.~(\ref{eq28}),
we find
\begin{equation}
\Phi_{\rm TE}(x)\approx -\int_x^{\infty}\!\!\!ydyr_{\rm TE}^2(x,y,0)e^{-y}.
\label{eq90}
\end{equation}
\noindent
Substituting here Eq.~(\ref{eq89}), one obtains
\begin{eqnarray}
&&
\Phi_{\rm TE}(x)\approx -Q_2^2x^2\int_x^{\infty}\!\!\!dy
\frac{y\,e^{-y}}{\left[Q_2x+2(y^2-x^2)\right]^2}
\nonumber \\
&&
\approx -\frac{Q_2^2x^2}{4}\int_x^{\infty}\!\!\!dy
\frac{e^{-y}}{y^3}=
\frac{Q_2^2x^2}{8}\left[{\rm Ei}(-x)-\frac{e^{-x}(1-x)}{x^2}\right]
\nonumber \\
&&~~
\approx -\frac{1}{8}Q_2^2\left[1-2x+x^2\ln x+O(x^2)\right].
\label{eq91}
\end{eqnarray}
\noindent
{}From this equation, the difference of our interest is given by
\begin{equation}
\Phi_{\rm TE}(\ri\tau t)-\Phi_{\rm TE}(-\ri\tau t)=
\ri\frac{Q_2^2}{2}\tau t.
\label{eq92}
\end{equation}

Comparing the difference in Eq.~(\ref{eq88}) with that in Eq.~(\ref{eq92}),
one finds that the latter is larger than the former by the factor
\begin{equation}
\frac{3Q_0^2Q_2^2}{8\pi^2Q_1}=\frac{24}{\pi^2}
\left(\frac{\alpha\sqrt{4\mu^2-\Delta^2}}{\vF\hbar\omega_c}\right)^3\gg 1.
\label{eq93}
\end{equation}
\noindent
Thus, one can approximately put
\begin{equation}
\Phi(\ri\tau t)-\Phi(-\ri\tau t)\approx
\Phi_{\rm TE}(\ri\tau t)-\Phi_{\rm TE}(-\ri\tau t).
\label{eq94}
\end{equation}

Finally, substituting Eqs.~(\ref{eq92}) and (\ref{eq94}) in Eq.~(\ref{eq29}),
one arrives at the result
\begin{equation}
\daT\cF\approx-\frac{k_BT}{16\pi a^2}Q_2^2\tau
\int_0^{\infty}\!\!\frac{t\,dt}{e^{2\pi t}-1}
=-\frac{4\alpha^2a(k_BT)^2(4\mu^2-\Delta^2)}{3\vF^2(\hbar c)^3}.
\label{eq95}
\end{equation}
\noindent
This result is obtained under a condition
$\sqrt{4\mu^2-\Delta^2}>\hbar\omega_c$ and, thus, $\mu\neq 0$.
However, $\Delta=0$ is allowed.

Now we consider the explicit contributions to the thermal correction in the
case $\Delta<2\mu$ starting with $\dbT\ocF$. We again use the condition
$\sqrt{4\mu^2-\Delta^2}>\hbar\omega_c$. Under this condition,
in accordance with Eq.~(\ref{eq80}), $\tp_{00,0}\yo=Q_0\neq 0$ and the
reflection coefficient $\rM(0,y,0)$ is given by the second expression in
Eq.~(\ref{eq83}) and, thus, is not equal to zero. Because of this, for
calculating the TM contribution to $\dbT\ocF$ one can use the term with
$l=0$ in Eq.~(\ref{eq31}).

The TE contribution to $\dbT\ocF$ is a different matter. Here, in accordance
to the second formula in Eq.~(\ref{eq80}), $\tp_0\yo= 0$ and, due to
Eq.~(\ref{eq89}), $\rE(0,y,0)=0$. Because of this, Eq.~(\ref{eq31}) is not
applicable in this case and one should calculate $\dbT\ocF_{\rm TE}$
using its definition as the term with $l=0$ in Eq.~(\ref{eq23}).
Taking into account that due to the equality $\rE(0,y,0)=0$ one has
$\rE(0,y,T)=\dT\rE(0,y,T)$, Eq.~(\ref{eq23}) leads to
\begin{eqnarray}
&&
\dbT\ocF_{\rm TE}(a,T)=\frac{k_BT}{16\pi a^2}\int_0^{\infty}\!\!\!\!
y\,dy
\ln\left\{1-\left[\dT\rE(0,y,T)\right]^2e^{-y}\right\}
\nonumber \\
&&~~
\approx -\frac{k_BT}{16\pi a^2}\int_0^{\infty}\!\!\!\!
ydy\left[\dT\rE(0,y,T)\right]^2e^{-y},
\label{eq96}
\end{eqnarray}
\noindent
where the last transformation is valid at sufficiently low $T$.

The thermal correction to the TM reflection coefficient in Eq.~(\ref{eq31}),
in accordance to Eqs.~(\ref{eq45}) and (\ref{eq80}) taken at $x=0$, is
given by
\begin{equation}
\dT\rM\oyt=\frac{2y\dT\tp_{00,0}\yt}{(Q_0+2y)^2}.
\label{eq97}
\end{equation}

For obtaining $\dT\rE$, Eq.~(\ref{eq45}) is not applicable, so that it is
found using Eq.~(\ref{eq4}) taken at $l=0$ with account of the equalities
$\tp_0=\dT\tp_0$ and $\rE\oyt=\dT\rE\oyt$
\begin{equation}
\dT\rE\oyt=-\frac{\dT\tp_{0}\yt}{\dT\tp_{0}\yt+2y^3}
\approx
-\frac{\dT\tp_{0}\yt}{2y^3}.
\label{eq98}
\end{equation}
\noindent
In the last transformation we have taken into account that the dominant
contribution to  Eq.~(\ref{eq96}) is given by $y\sim 1$ and that
$\dT\tp_0$ goes to zero with vanishing $T$.

In the case $\Delta<2\mu$ under consideration now, the quantities
$\dT\tp_{00,0}$ and $\dT\tp_0$, entering Eqs.~(\ref{eq97}) and (\ref{eq98}),
can be found from Eqs.~(\ref{eq10}) and (\ref{eq16a})
\begin{widetext}
\begin{eqnarray}
&&
\dT\tp_{00,0}\yt=\frac{4\alpha D}{\vF^2}\left[
\int_1^{\infty}\!\!dt\left(e^{\frac{t\Delta-2\mu}{2k_BT}}+1\right)^{-1}
\!\!X_{00,0}(t,y,D)
-\int_1^{2\mu/\Delta}\!\!\!\!dtX_{00,0}(t,y,D) \right],
\nonumber \\[-1mm]
&&\label{eq99}\\[-1mm]
&&
\dT\tp_{0}\yt=
\frac{4\alpha D}{\vF^2}\left[
\int_1^{\infty}\!\!dt\left(e^{\frac{t\Delta-2\mu}{2k_BT}}+1\right)^{-1}
\!\!X_{0}(t,y,D)
-\int_1^{2\mu/\Delta}\!\!\!\!\!dtX_{0}(t,y,D) \right].
\nonumber
\end{eqnarray}
\end{widetext}
\noindent
Here, similar to Eqs.~(\ref{eq46}) and (\ref{eq47}), we have omitted the
first contribution to Eq.~(\ref{eq11})  leading to an additional exponentially
small factor.

The quantities $X_{00,0}$ and $X_0$ in Eq.~(\ref{eq99}) are defined by
Eq.~(\ref{eq12}) where one should put $l=0$
\begin{eqnarray}
&&
X_{00,0}(t,y,D)=1+\frac{1}{\vF y} {\rm Re}
\frac{D^2t^2-\vF^2y^2}{\sqrt{\vF^2y^2-D^2t^2+D^2}},
\nonumber \\
&&
X_{0}(t,y,D)={\vF y}D^2 {\rm Re}\frac{t^2-1}{\sqrt{\vF^2y^2-D^2t^2+D^2}}.
\label{eq100}
\end{eqnarray}
\noindent
Note that here the real part is not equal to zero only for
$t\leqslant f(y,D)$,
where $f(y,D)$ is defined in Eq.~(\ref{eq48}). It is easily seen that
$f(y,D)<2\mu/\Delta$ [the upper integration limit in the second contributions
in Eq.~(\ref{eq99})] if $y$ satisfies the inequality
\begin{equation}
y<\frac{\sqrt{4\mu^2-\Delta^2}}{\vF\hbar\omega_c}.
\label{eq101}
\end{equation}

Under the condition $\sqrt{4\mu^2-\Delta^2}>\hbar\omega_c$, accepted above,
this inequality is satisfied with large safety margin over the entire range of
$y$ giving the major contribution to Eqs.~(\ref{eq31}) and (\ref{eq96}).
Because of this, the upper integration limits of the integrals with respect
to $t$ in Eq.~(\ref{eq99}), containing the real parts indicated  in
Eq.~(\ref{eq100}), should be replaced with $f(y,D)$. Taking into account also
that $D>1$, i.e., $D\gg\vF y$, and $t^2-1<\vF^2y^2/D^2$ over the entire
domain of integration, from  Eqs.~(\ref{eq99}) and (\ref{eq100}) in the
asymptotic limit $k_BT\ll 2\mu-\Delta$ one obtains
\begin{eqnarray}
&&
\dT\tp_{00,0}\yt=\frac{4\alpha D}{\vF^2}
\times\left[
\int_1^{\infty}\!\!dt\left(e^{\frac{t\Delta-2\mu}{2k_BT}}+1\right)^{-1}
-\int_1^{2\mu/\Delta}\!\!\!dt\right]
+\frac{4\alpha D^3}{\vF^3y}Y\yt,
\nonumber \\
&&
\dT\tp_{0}\yt=\frac{4\alpha\vF y^3}{D}Y\yt,
\label{eq102}
\end{eqnarray}
\noindent
where the following notation is introduced
\begin{equation}
Y\yt\equiv\int_1^{f(y,D)}\!\!dt
\left[\left(e^{\frac{t\Delta-2\mu}{2k_BT}}+1\right)^{-1}-1\right]
\frac{1}{\left[\vF^2y^2-D^2(t^2-1)\right]^{1/2}}. \label{eq103}
\end{equation}

The first contribution to $\dT\tp_{00,0}$ in Eq.~(\ref{eq102}) is
easily calculated
\begin{eqnarray}
&&~
\frac{4\alpha D}{\vF^2}\left[
\int_1^{\infty}\!\!dt\left(e^{\frac{t\Delta-2\mu}{2k_BT}}+1\right)^{-1}
-\int_1^{2\mu/\Delta}\!\!\!dt\right]
=\frac{8\alpha}{\vF^2\hbar\omega_c}\left[k_BT
\ln\frac{\left(1+e^{\frac{\Delta-2\mu}{2k_BT}}\right)
\left(1+e^{\frac{\mu}{k_BT}}\right)}{1+e^{-\frac{\mu}{k_BT}}}
-\mu\right]
\nonumber \\
&&~~
\approx\frac{8\alpha}{\vF^2\hbar\omega_c}\meE.
\label{eq104}
\end{eqnarray}

The low-temperature behavior of the integral $Y$ defined in Eq.~(\ref{eq103})
is found in the Appendix. According to Eq.~(\ref{A9}) one has
\begin{equation}
Y\yt\approx -\frac{\vF y}{D^2}\,\meE.
\label{eq105}
\end{equation}

Substituting  Eqs.~(\ref{eq104}) and (\ref{eq105}) in Eq.~(\ref{eq102}),
one obtains
\begin{eqnarray}
&&
\dT\tp_{00,0}\yt\approx\frac{4\alpha}{\vF^2}
\left(\frac{2k_BT}{\hbar c}-D\right)\meE
\sim
-\frac{\alpha\Delta}{\hbar\omega_c}\meE,
\nonumber \\
&&
\dT\tp_{0}\yt\approx
-\frac{4\alpha\vF^2y^4}{D^3}\meE.
\label{eq108}
\end{eqnarray}
\noindent

We note that according to  Eq.~(\ref{eq96}) $\dbT\ocF_{\rm TE}$ is of the
order of $(\dT\rE)^2$, i.e., $\sim(\dT\tp_0)^2\sim \exp[-2(2\mu-\Delta)/(2k_BT)]$
and, thus, contains an additional exponentially small factor.
Because of this, we have
\begin{equation}
\dbT\cF\approx\dbT\ocF_{\rm TM}(a,T).
\label{eq109}
\end{equation}

Substituting Eqs.~(\ref{eq83}), (\ref{eq97}), and the first equality
in Eq.~(\ref{eq108}) in  the TM term of Eq.~(\ref{eq31}) with $l=0$,
one finally finds
\begin{equation}
\dbT\ocF_{\rm TM}(a,T)\sim\frac{k_BTQ_0\alpha\Delta}{a^2\hbar\omega_c}
\,\meE
\int_0^{\infty}\!\!\!\!y^2dy\frac{e^{-y}}{(Q_0+2y)^3-Q_0(Q_0+2y)e^{-y}}.
\label{eq110}
\end{equation}
\noindent
Taking into account Eq.~(\ref{eq109}), the convergence of the integral which
is of the order of $Q_0^{-3}$, and substituting the definition of $Q_0$
given  in Eq.~(\ref{eq81}), the up to an order of magnitude behavior of
$\dbT\ocF$ at low temperature is
\begin{equation}
\dbT\cF\sim \frac{k_BT\hbar c\Delta}{\alpha a^3\mu^2}\,\meE.
\label{eq111}
\end{equation}

We recall that this asymptotic behavior is derived under the conditions $D>1$,
i.e., $\Delta>\hbar\omega_c$ and  $\sqrt{4\mu^2-\Delta^2}>\hbar\omega_c$
which are satisfied at sufficiently large separations between graphene sheets
with nonzero $\Delta$ and $\mu$.

It only remains to find the low-temperature behavior of the last contribution
to the thermal correction $\dcT\ocF$. We note that for
$l\geqslant 1$ both the quantities $\tp_{00,l}\yo\neq 0$ and
$\tp_{l}\yo\neq 0$ so that $\dcT\ocF$ is given by sum of all terms with
$l\geqslant 1$  in Eq.~(\ref{eq31}). In doing so, it will suffice to preserve
the dependence on $\tau$ ($\zeta_l=\tau l$) only in the lower integration limits
of all integrals in Eq.~(\ref{eq31}) and substitute the integrands in the lowest
perturbation order in $\tau$. For the TM mode, this means that one should use in
Eq.~(\ref{eq31}) the second line in Eq.~(\ref{eq83}), Eq.~(\ref{eq97}), and
the first line in Eq.~(\ref{eq108}). For the TE mode, according to
Eq.~(\ref{eq89}), $\rE(0,y,0)=0$. Because of this, $\rE\ozy$ should be taken
in the first perturbation order in $\tau$ as given by Eq.~(\ref{eq89}),
whereas the thermal
correction to the TE reflection coefficient is given by Eq.~(\ref{eq98})
and by the second line in Eq.~(\ref{eq108}).

As a result, for the contribution of the TM mode one obtains
\begin{equation}
\dcT\ocF_{\rm TM}(a,T)\sim \frac{k_BT\hbar c\Delta}{\alpha\mu^2a^3}\meE
\sum_{l=1}^{\infty}\int_{\zeta_l}^{\infty}\!\!\frac{y^2dy}{e^{y}-1},
\label{eq112}
\end{equation}
\noindent
where we have used that $y$ giving the major contribution to the integral
satisfies the condition $y\ll Q_0$.

For the sum of integrals in Eq.~(\ref{eq112}) we have
\begin{eqnarray}
&&
\sum_{l=1}^{\infty}\int_{\zeta_l}^{\infty}\!\!\frac{y^2dy}{e^{y}-1}
=\sum_{n=1}^{\infty}\frac{1}{n^3}\int_{n\zeta_l}^{\infty}\!\!\!\!
dxx^2e^{-x}
=\sum_{n=1}^{\infty}\left[\frac{2}{n^3}\frac{1}{e^{\tau n}-1}+
\frac{2\tau}{n^2}\frac{e^{\tau n}}{(e^{\tau n}-1)^2}+
\frac{\tau^2}{n}\frac{e^{\tau n}(1+e^{\tau n})}{(e^{\tau n}-1)^3}\right]
\nonumber \\
&&
\sim \frac{1}{\tau}\sum_{n=1}^{\infty}\left[\frac{2}{n^4}+\frac{2}{n^4}
+\frac{1}{n^4}\right]
\sim \frac{1}{\tau}.
\label{eq113}
\end{eqnarray}
Substituting this to Eq.~(\ref{eq112}), we arrive at
\begin{equation}
\dcT\ocF_{\rm TM}(a,T)\sim
\frac{(\hbar c)^2\Delta}{\alpha\mu^2a^4}\meE  .
\label{eq114}
\end{equation}

The contribution of the TE mode is obtained by substituting  Eqs.~(\ref{eq89}),
(\ref{eq98}), and (\ref{eq108}) in Eq.~(\ref{eq31}) at low $T$
\begin{equation}
\dcT\ocF_{\rm TE}(a,T)\sim \frac{\alpha k_BT}{a^2}
\frac{Q_2}{D^3}\meE\tau
\sum_{l=1}^{\infty}l\int_{\zeta_l}^{\infty}\!\!dy e^{-y}
\sim \frac{\alpha^2\sqrt{4\mu^2-\Delta^2}(\hbar c)^3}{\Delta^3a^5}\,\meE.     \label{eq115}
\end{equation}

It is easily seen that the quantities in Eqs.~(\ref{eq114}) and
(\ref{eq115}) can be of the same order of magnitude. Thus, for the total
contribution $\dcT\ocF$ we obtain
\begin{equation}
\dcT\cF\sim
\frac{(\hbar c)^2}{a^4}\left(\frac{\Delta}{\alpha\mu^2}
+\frac{\alpha^2\hbar c\sqrt{4\mu^2-\Delta^2}}{a\Delta^3}
\right)\meE     .
\label{eq116}
\end{equation}
\noindent
This result is derived for $\mu>0$ and $\Delta>0$.

{}From   Eqs.~(\ref{eq95}), (\ref{eq111}), and (\ref{eq116}), one
concludes that the main term in the low-temperature behavior of the
Casimir free energy for graphene with $\Delta<2\mu$ is determined by the
TE mode in the implicit contribution given by  Eq.~(\ref{eq95}).
Substituting Eq.~(\ref{eq95}) in Eq.~(\ref{eq70}) one arrives at the
Casimir entropy at low temperature
\begin{equation}
S(a,T)\sim \frac{\alpha^2a(4\mu^2-\Delta^2)k_B^2T}{(\hbar c)^3}.
\label{eq117}
\end{equation}
\noindent
In the limit of vanishing temperature, the Casimir entropy (\ref{eq117})
goes to zero in agreement with the Nernst heat theorem.

The results of this section were derived under the conditions
\begin{equation}
k_BT\ll\frac{\hbar v_F}{2a}\ll\frac{\hbar c}{2a}<\Delta,
\quad
k_BT\ll 2\mu-\Delta.
\label{eq115a}
\end{equation}
\noindent
Thus, although the first two expansion parameters in Eq.~(\ref{eq73b})
remain the same, the third one is replaced with
\begin{equation}
e^{-\frac{2\mu-\Delta}{2k_BT}}\ll 1.
\label{eq115b}
\end{equation}
\noindent
One more condition used in the derivation of expressions (\ref{eq80})
for the polarization tensor is
\begin{equation}
\frac{\hbar c}{2a}<\sqrt{4\mu^2-\Delta^2}.
\label{eq115c}
\end{equation}
\noindent
These application conditions are discussed in Sec.~VII.

\section{CONCLUSIONS AND DISCUSSION}

In this paper, we have found the low-temperature behavior of the
Casimir free energy and entropy of two real graphene sheets possessing
the nonzero energy gap and chemical potential. This problem is solved
analytically in the framework of the Dirac model. The response of
graphene to the electromagnetic field is described on the basis of
first principles of thermal quantum field theory by means of the
polarization tensor in (2+1)-dimensional space-time. The thermal
correction to the Casimir energy of two parallel graphene sheets at
zero temperature is presented as a sum of two contributions. The
first of them, called implicit, contains the polarization tensor at
zero temperature, and the dependence of this contribution on
temperature is determined by a summation over the Matsubara
frequencies. The temperature dependence of the second contribution,
called explicit, is determined by the thermal correction to the
polarization tensor. The low-temperature behaviors of both
contributions were found for different relationships between the
energy gap and chemical potential of graphene sheets, i.e., for
$\Delta>2\mu$, $\Delta=2\mu$, and $\Delta<2\mu$, and turned out
to be essentially different.

According to the results of Sec.~IV, which are repeated here by
presenting only the dimensional quantities, the low-temperature
behavior of the Casimir free energy and entropy for graphene sheets
with $\Delta>2\mu$ is eventually determined by the TE mode in an
implicit contribution to the thermal correction
\begin{equation}
\dT\cF\sim -\frac{(k_BT)^5}{(\hbar c)^2\Delta^2},\quad
S(a,T)\sim\frac{k_B^5T^4}{(\hbar c)^2\Delta^2},
\label{eq118}
\end{equation}
\noindent
and it does not depend on the chemical potential.

In Sec.~V it is shown that for graphene sheets with $\Delta=2\mu$
the eventual low-temperature behavior of the Casimir free energy
and entropy is determined by the TM mode in an explicit contribution
to the thermal correction
\begin{equation}
\dT\cF\sim -\frac{k_BT}{a^2}, \quad
S(a,T)\sim \frac{k_B}{a^2},
\label{eq119}
\end{equation}
\noindent

Finally, as shown in Sec.~VI, for the case $\Delta<2\mu$ the
low-temperature behavior of the Casimir free energy and entropy is
governed by the TE mode in an implicit contribution to the thermal
correction given by
\begin{equation}
\dT\cF\sim-\frac{a(4\mu^2-\Delta^2)(k_BT)^2}{(\hbar c)^3},
\quad
S(a,T)\sim\frac{a(4\mu^2-\Delta^2)k_B^2T}{(\hbar c)^3}.
\label{eq120}
\end{equation}
\noindent

It is interesting to compare these results with the case of a
pristine graphene with $\Delta=\mu=0$ where \cite{53}
\begin{equation}
\dT\cF\sim \frac{(k_BT)^3}{(\hbar c)^2}\ln\frac{ak_BT}{\hbar c},
\quad
S(a,T)\sim-k_B\frac{(k_BT)^2}{(\hbar c)^2}\ln\frac{ak_BT}{\hbar c}.
\label{eq121}
\end{equation}
\noindent

As is seen from the comparison of Eqs.~(\ref{eq118})--(\ref{eq120})
with Eq.~(\ref{eq121}), for real graphene sheets there is a nontrivial
interplay between the values of $\Delta$ and $\mu$ which leads to
different behaviors of the Casimir energy and entropy with vanishing
temperature, especially in the case $\Delta<2\mu$ where the
polarization tensor at $T=0$ depends on $\mu$.

{}From Eqs.~(\ref{eq118}) and (\ref{eq120}) one concludes that the
Casimir entropy is positive and vanishes with vanishing temperature,
i.e., for graphene with $\Delta>2\mu$ and $\Delta<2\mu$ the Nernst
heat theorem is satisfied and, thus, the Lifshitz theory of the
Casimir interaction is consistent with the requirements of
thermodynamics (the same holds for a pristine graphene). According
to Eq.~(\ref{eq119}), this is, however, not so for graphene with
$\Delta=2\mu\neq 0$ where the Casimir entropy at zero temperature
is not equal to zero and its value depends on the parameter of a
system (volume). As discussed in Sec.~V, however, this anomaly is not
relevant to any physical situation because for real graphene samples
the exact equality $\Delta=2\mu$ is not realizable. We note that the
real part of the electrical conductivity of graphene as a function of
frequency also experiences a qualitative change when the energy gap
$\Delta$ decreases from $\Delta>2\mu$ to $\Delta<2\mu$ \cite{37}.

It should be noted that the asymptotic expressions (\ref{eq118}) and
(\ref{eq120}) are not applicable to graphene sheets with too small
values of $\Delta-2\mu$ and $2\mu-\Delta$, respectively.
The point is that if the values of $\Delta$ and $2\mu$ are too close
to each other the exponentially small parameters in Eqs.~(\ref{eq73b})
and (\ref{eq115b}) lose their meaning and cannot be used.
Taking into account that the polarization tensor is a continuous function
of $\Delta$ at the point $\Delta=2\mu$, the possibility exists that an
apparent discontinuity of the obtained asymptotic formulas at $\Delta=2\mu$
may be an artifact of the expansion in small parameters at the crossover
region. For a comprehensive resolution of this question, it would be
desirable to find the more exact asymptotic expressions applicable for the
values of $2\mu$   arbitrarily close to $\Delta$ from the left and from
the right. In future it is also interesting to investigate the case of two
dissimilar graphene sheets with different values of the energy gap and
chemical potential. The configuration of a graphene sheet interacting
with an ideal metal plane (it has been known that for two ideal metal planes
the Casimir entropy satisfies the Nernst heat theorem \cite{88}) or
a plate made of conventional metallic or dielectric materials.

According to Sec.~I, theoretical description of the Lifshitz theory
using the polarization tensor of graphene \cite{52} is in good
agreement with the experiment on measuring the Casimir interaction in
graphene system \cite{51}. Taking into consideration that the
polarization tensor of graphene results in two spatially nonlocal,
complex dielectric permittivities (the longitudinal one and the
transverse one \cite{27}), it may be suggested that a more fundamental
theoretical description of the dielectric response of metals admits
a similar approach. In application to metals, the nonlocal dielectric
permittivities of this kind could lead to almost the same results,
as the dissipative Drude model, for the propagating waves on the mass
shell, but deviate from them significantly for the evanescent fields
off the mass shell (in contrast to the nonlocal dielectric functions
describing the anomalous skin effect \cite{63}). In such a manner
graphene might point the way for resolution of the Casimir puzzle
which remains unresolved for already 20 years.

\section*{Acknowledgments}
The work of G.L.K. and V.M.M. was partially supported by the Peter
the Great Saint Petersburg Polytechnic University in the framework
of the Program ``5--100--2020". The work of V.M.M. was partially funded
by the Russian Foundation for Basic Research, Grant No. 19-02-00453 A.
His work was also partially supported by the Russian Government Program
of Competitive Growth of Kazan Federal University.
\appendix
\section{}
\setcounter{equation}{0}
\renewcommand{\theequation}{A\arabic{equation}}

Here, we derive the low-temperature behavior of two integrals used in
the main text.
We begin with the integral $I_{00,0}^{(2)}$ defined in Eqs.~(\ref{eq47})
and (\ref{eq48}). To calculate the quantity $I_{00,0}^{(2)}$  in the case
$\Delta>2\mu$ we introduce the integration variable $v=t-1$
and obtain
\begin{equation}
I_{00,0}^{(2)}\approx \deE\int_0^{f(y,D)-1}\!\!\!dv e^{-v\frac{\Delta}{2k_BT}}
\frac{D^2(v+1)^2}{[\vF^2y^2-D^2v(v+2)]^{1/2}},
\label{A1}
\end{equation}
\noindent
where we have omitted the negligibly small quantity $\vF^2y^2$ taking into
account that the dominant contribution to Eq.~(\ref{eq31}) is given
by $y\sim 1$.
Under this condition $f(1,D)-1\ll 1$ and, thus, $v\ll 1$.
Then the asymptotic behavior of Eq.~(\ref{A1}) at low $T$ can be estimated as
\begin{eqnarray}
&&
I_{00,0}^{(2)}\sim \deE\int_0^{f(1,D)-1}\!\!\!dv e^{-v\frac{\Delta}{2k_BT}}
\frac{D^2}{(\vF^2-2D^2v)^{1/2}}
\nonumber \\[-1mm]
&&\label{A2} \\[-1mm]
&&~~
=D\frac{2k_BT}{\Delta}\deE\int_0^{U(D,T)}du
\frac{e^{-u}}{\left(\frac{\vF^2}{D^2}-4\frac{k_BT}{\Delta}u\right)^{1/2}},
\nonumber
\end{eqnarray}
\noindent
where $u=v\Delta/(2k_BT)$ is the integration variable introduced in place of $v$,
and $U(D,T)\equiv\Delta[f(1,D)-1]/(2k_BT)$. In view of the fact that
$4k_BTu/\Delta$ goes to zero when $T$ vanishes and the main contribution to the
integral is given by $u\sim 1$, we find
\begin{equation}
I_{00,0}^{(2)}\sim \frac{D^2}{\vF}\frac{k_BT}{\Delta}\deE
\int_0^{\infty}\!\!\!du e^{-u}= \frac{k_BT}{\vF}
\frac{\Delta}{(\hbar \omega_c)^2}
\deE.
\label{A3}
\end{equation}

Now we consider the same integral but for graphene with $\Delta=2\mu$.
For this purpose, we again begin from Eq.~(\ref{eq47}), where now $\Delta=2\mu$,
and substitute there the identity
\begin{equation}
\left[e^{\frac{(t-1)\Delta}{2k_BT}}+1\right]^{-1}=
\sum_{n=1}^{\infty}(-1)^{n-1}e^{-n(t-1)\frac{\Delta}{2k_BT}}.
\label{A4}
\end{equation}
\noindent
Then, after introducing the integration variable $v=t-1$, one obtains instead
of Eq.~({\ref{A2})
\begin{eqnarray}
&&
I_{00,0}^{(2)}\sim \sum_{n=1}^{\infty}(-1)^{n-1}\!\!
\int_0^{f(1,D)-1}\!\!\!dv e^{-nv\frac{\Delta}{2k_BT}}
\frac{D^2}{(\vF^2-2D^2v)^{1/2}}
\nonumber \\[-1mm]
&&\label{A5} \\[-1mm]
&&
=D\frac{2k_BT}{\Delta}\sum_{n=1}^{\infty}\frac{(-1)^{n-1}}{n}\!\!
\int_0^{nU(D,T)}du
\frac{e^{-u}}{\left(\frac{\vF^2}{D^2}-4\frac{k_BT}{n\Delta}u\right)^{1/2}},
\nonumber
\end{eqnarray}
\noindent
where $u=nv\Delta/(2k_BT)$. For arbitrarily small $T$ this equation can be
rearranged as
\begin{equation}
I_{00,0}^{(2)}\sim \frac{D^2}{\vF}\frac{k_BT}{\Delta}
\sum_{n=1}^{\infty}\frac{(-1)^{n-1}}{n}
\int_0^{\infty}\!\!\!du e^{-u}
=\frac{k_BT}{\vF}
\frac{\Delta}{(\hbar \omega_c)^2}
\ln 2.
\label{A6}
\end{equation}

Now we find the low-temperature behavior of the integral $Y$ defined in
Eq.~(\ref{eq103}). The power of exponent in Eq.~(\ref{eq103}) is negative
over the entire integration range. Because of this, one can use the
following expansion
\begin{widetext}
\begin{eqnarray}
&&
Y\yt=\int_1^{f(y,D)}\!\!dt\left[\sum_{n=1}^{\infty}(-1)^{n-1}
e^{\frac{t\Delta-2\mu}{2k_BT}(n-1)}-1\right]
\frac{1}{\left[\vF^2y^2-D^2(t^2-1)\right]^{1/2}}
\nonumber \\
&&~~~~
=-\sum_{k=1}^{\infty}(-1)^{k-1}\int_1^{f(y,D)}\!\!dt
e^{\frac{t\Delta-2\mu}{2k_BT}k}
\frac{1}{\left[\vF^2y^2-D^2(t^2-1)\right]^{1/2}}.
\label{A7}
\end{eqnarray}
\end{widetext}

Now we replace the integration variable $t$ with $v=t-1$ and take into
account that for $y\sim 1$
\begin{equation}
f(y,D)-1\approx\frac{\vF^2y^2}{2D^2}\ll 1.
\label{A8}
\end{equation}
\noindent
For this reason, one can neglect by $v$ as compared to unity in the power
of exponent and also in the denominator of Eq.~(\ref{A7}). In the sum, we
can restrict ourselves by only the first term because all other terms
contain additional exponentially small factors as compared with it.
The result is
\begin{equation}
Y\yt\approx -\meE\!\!\int_0^{f(y,D)-1}\!\!
\frac{dv}{\left(\vF^2y^2-2D^2v\right)^{1/2}}
\approx -\frac{\vF y}{D^2}\meE,
\label{A9}
\end{equation}
\noindent
where we have used the condition (\ref{A8}).

\end{document}